\documentclass[12pt,preprint]{aastex}

\shorttitle{Discovery of the Ammonium Ion}
\shortauthors{Cernicharo et al.}

\begin{document}


\title{Detection of the Ammonium Ion in Space
\thanks{This work was based on observations carried out with the 
IRAM 30-meter telescope. IRAM is supported by INSU/CNRS (France), 
MPG (Germany) and IGN (Spain)}}

\author{J. Cernicharo \& B. Tercero}
\affil{Deparment of Astrophysics, CAB. INTA-CSIC. Crta Torrej\'on-Ajalvir Km\,4. 28850 Torrej\'on
de Ardoz, Madrid, Spain}
\email{jcernicharo@cab.inta-csic.es}

\and
\author{A. Fuente}
\affil{Observatorio Astron\'omico Nacional, Apdo. 112, 28803, Alcal\'a de Henares, Spain}

\and
\author{J.L. Domenech, M. Cueto, E. Carrasco, V.J. Herrero \& I. Tanarro}
\affil{Instituto de Estructura de la Materia, IEM-CSIC, Serrano 123,
28006 Madrid, Spain}

\and
\author{N. Marcelino}
\affil{NRAO, 520 Edgemont Road, Charlottesville, VA 22902, USA}

\and
\author{E. Roueff}
\affil{Luth, Observatoire de Paris, CNRS UMR8102, Place J. Janssen F-92190 Meudon, France}

\and
\author{M. Gerin}
\affil{LERMA, Observatoire de Paris, CNRS UMR8112 and Ecole Normale Superieure, 61 Avenue de l'Observatoire, F-75014 Paris, France}

\and
\author{J. Pearson}
\affil{Jet Propulsion Laboratory, 4800 Oak Grove Drive, MC 168-314, Pasadena, CA 91109, USA}

\begin{abstract}
We report on the detection of a narrow feature at 262816.73\,MHz towards Orion and
the cold prestellar core B1-bS, that we
attribute to the 1$_0$-0$_0$ line of the deuterated Ammonium ion, 
NH$_3$D$^+$.
The observations were performed with the IRAM 30m radio telescope.
The carrier has to be a light molecular species as it is the only feature
detected over 3.6 GHz of bandwidth. The hyperfine structure is not resolved
indicating a very low value for the electric quadrupolar coupling constant of Nitrogen which
is expected for NH$_3$D$^+$ as the electric field over the N nucleus is practically
zero. Moreover, the feature is right at the predicted frequency
for the 1$_0$-0$_0$ transition of the Ammonium ion, 262817$\pm$6 MHz (3$\sigma$), 
using rotational constants derived from new infrared data obtained in our laboratory in Madrid.
The estimated column density is (1.1$\pm$0.2$)\times$10$^{12}$ cm$^{-2}$. 
Assuming a deuterium enhancement similar to that of NH$_2$D, 
we derive $N$(NH$_4^+$)$\simeq$ 2.6$\times$10$^{13}$ cm$^{-2}$, i.e., an abundance
for Ammonium of a few 10$^{-11}$.

\end{abstract}


\section{Introduction}
Two of the major nitrogen bearing molecules in the ISM are
N$_2$ and NH$_3$, which are predicted to be present in many different
media (see, e.g., \citealt{Nejad1990, Aikawa2008, Harada2010}). 
Whereas N$_2$ is expected to be largely in the gas phase,
NH$_3$ should be mostly frozen onto the surface of dust grains for
temperatures lower than 100 K.
Diazenylium (N$_2$H$^+$) and Ammonium (NH$_4^+$), the protonated
ions of N$_2$ and NH$_3$, can also provide crucial information for the
understanding of interstellar chemistry. The N$_2$H$^+$ ion is usually
taken as a tracer for the apolar N$_2$ molecule, which is practically
impossible to observe directly using conventional spectroscopic
techniques. Detection of diazenylium has been reported from
many different sources \citep{Turner1974,Green1974,Fuente1993,Daniel2007}. 
The ion is assumed to be formed almost exclusively through the reaction of N$_2$ molecules with H$_3^+$
and to be destroyed mostly in collisions with electrons or with CO
molecules.
The existence of interstellar NH$_4^+$ is also predicted
in various astrochemical models. In particular, in hot cores \citep{Aikawa2011}
and in other high temperature environments like active galactic
nuclei \citep{Harada2010}, grain evaporation leads to a large increase in the gas phase
concentration of NH$_3$. Ammonia has a high proton
affinity and its protonated derivative, 
NH$_4^+$, is predicted to become one of the dominant ions \citep{Aikawa2008,Aikawa2011}. 
Ammonium is also the dominant ion in many cold laboratory plasmas containing hydrogen and nitrogen 
\citep{Carrasco2011,Carrasco2013}.
However, the symmetric character of NH$_4^+$ precludes its observation through radioastronomic methods, and NH$_4^+$ has never been 
detected in the ISM. Therefore, the deuterated variant NH$_3$D$^+$, possessing a small permanent dipole moment, is more suitable for 
astronomical searches. 

Deuterated Ammonia NH$_2$D was detected in Orion by \citet{Rodriguez1978} and in SgrB2
by \citet{Turner1978}. A detailed analysis of the NH$_2$D/NH$_3$ abundance ratio in Orion
was first provided by \citet{Walmsley1987} who obtained a value of 0.003. Deuterated Ammonia
is produced through the dissociative recombination of NH$_3$D$^+$ with electrons. 
NHD$_2$, and ND$_3$ have been
also detected towards a large variety of molecular clouds harbouring a large range of kinetic
temperatures (\citealt{Saito2000,Roueff2000,Lis2002,
Roueff2005,Lis2006}). Hence, the precursor molecule NH$_3$D$^+$, and even more
deuterated isotopologues of the Ammonium ion, could be present in hot and cold molecular clouds. 

In this Letter we report on the first detection in space of the singly deuterated Ammonium ion
towards the sources Orion-IRc2 and the cold prestellar core B1-bS, a source where the
whole family of $^{15}$N and deuterated isotopologues of Ammonia and NNH$^+$ have been detected
(\citealt{Lis2002,Lis2006,Saito2000}; Daniel et al., in preparation).

\section{Observations}
The observations presented in this paper were motivated by the
analysis of the Orion line survey carried out with the 30m IRAM
telescope \citep{Tercero2010, Tercero2011}. During the interpretation of the
line survey we had to deal with more than 8000 unidentified features. Around 4400
of them have been successfully assigned to several
isotopes of CH$_3$CH$_2$CN, CH$_2$CHCN, CH$_3$OCOH, their vibrational levels
and those of all abundant molecules, and to the recently detected
methyl acetate and the gauche conformer of ethyl formate in this source
(\citealt{Demyk2007, Carvajal2009, Margules2009, Margules2010, Tercero2010, Tercero2011, Tercero2012, 
Tercero2013, Motiyenko2012, Daly2013}; L\'opez et al. in preparation). Among the 
strongest unidentified
features we found one at 262816.7\,MHz for $v$$_{LSR}$=9\,km\,s$^{-1}$,
(see Figure 1) that was very close to the
predicted frequency for the $J_K$=1$_0$-0$_0$ line of NH$_3$D$^+$. This deuterated species
of the Ammonium ion was observed in the laboratory by \citet{Nakanaga1986} through its
$\nu_4$ vibrational mode from which rotational constants for the ground state were derived.
The molecule was incorporated some time ago to the MADEX code \citep{Cernicharo2012}. The
calculated frequency for the $J_K$=1$_0$-0$_0$ transition of NH$_3$D$^+$ from these rotational
constants is 262807\,MHz with a $\pm$3$\sigma$ uncertainty of $\pm$9\,MHz.
The spectrum shown in Figure 1 shows a line 10\,MHz away
from the predicted one with a linewidth similar to those of the NH$_2$D lines observed in the
same frequency survey. 
The observed feature, labelled U262816.7 in Figure 1, is practically
free of blending from other species.
None of the rotational transitions of the recently characterized
isotopologues and vibrationally excited states of methyl/vinyl/ethyl cyanide, methyl formate, formamide,
or from the abundant species found in Orion \citep{Tercero2010}
matches the frequency of U262816.7 except
a possible very weak contribution from $^{13}$CH$_3$OCOH in its first torsional state.
Nevertheless, having still 3600 unidentified features in Orion, claiming a possible identification with
NH$_3$D$^+$ will be extremely risky and speculative without a more precise frequency measurement
in the laboratory (see Section 4).

\begin{figure}
\includegraphics[angle=0,scale=.5]{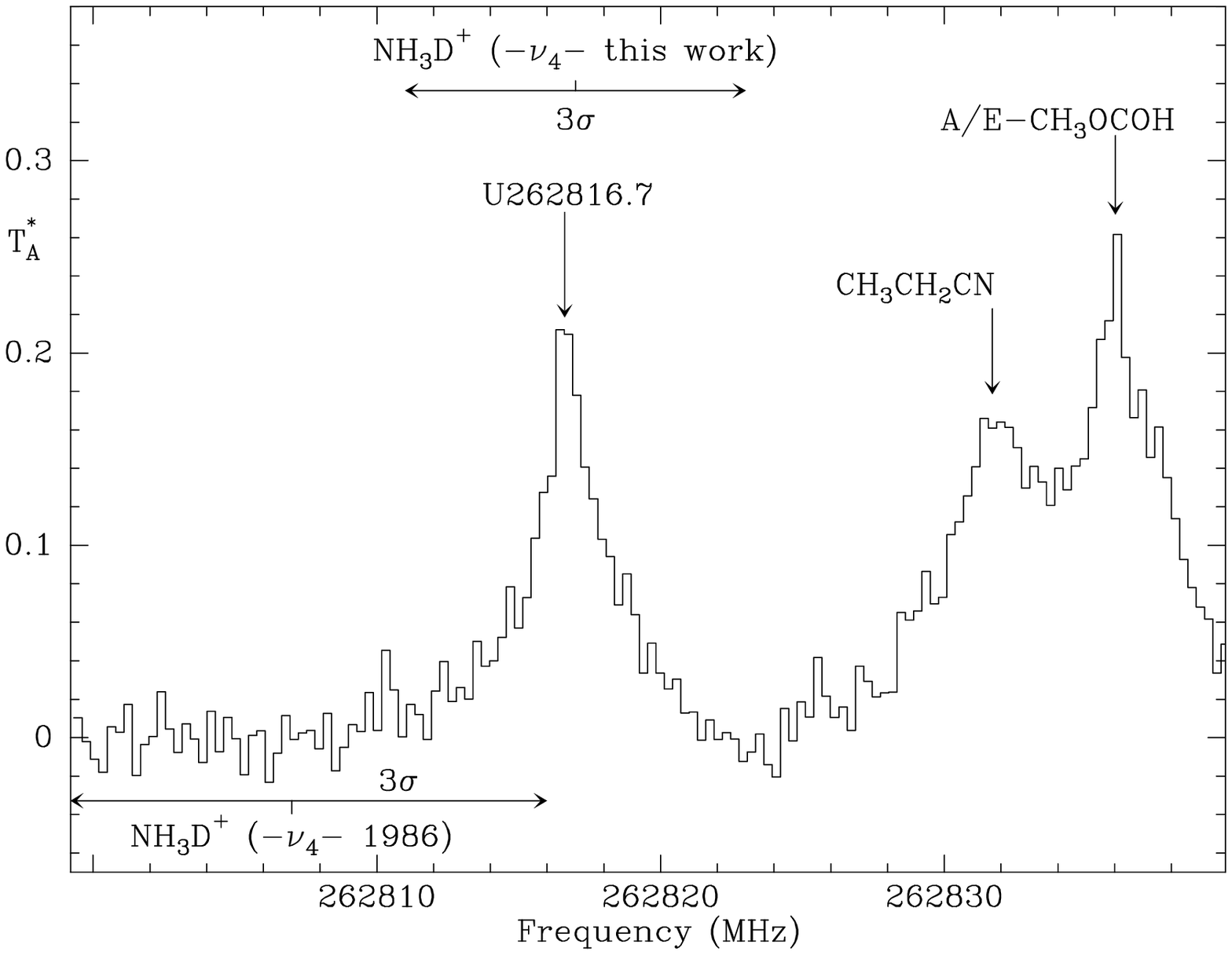}
\caption{Observed spectrum towards Orion-IRc2 around the expected
frequency of the $J_K$=1$_0$-0$_0$ line of the deuterated ammonium ion.
Identification of all other lines in this frequency range is indicated. The predicted
frequency and its 3$\sigma$ error bar is indicated at the bottom for the \citet{Nakanaga1986}
measurements and at the top for our new laboratory data (Domenech et al., 2013; see accompanying
Letter).}
\end{figure}

Prompted by the observed line in Orion we started a search for a similar feature 
towards the cold prestellar cloud B1-bS.
Complex organic molecules have been recently found in this cold core \citep{Marcelino2009,Marcelino2010,
Cernicharo2012b}. This source also shows an impressive enhancement in the deuteration
of abundant molecules such as H$_2$CO, H$_2$CS, and NH$_3$ \citep{Lis2002,Lis2006,Marcelino2005,
Saito2000}. With a kinetic temperature of $\simeq$12\,K the
expected density of lines at 262.8 GHz is rather low. Previous
observations at 3 and 2 millimeter wavelengths 
indicate that only some diatomic and triatomic species will have significant 
emission at these high frequencies \citep{Marcelino2005, Marcelino2009, Marcelino2010,
Cernicharo2012b}.

The observations were performed 
at the IRAM 30m telescope at Pico Veleta (Spain)
using the 1\,mm Eight MIxer Receivers (EMIR). The backend was
a fast Fourier Transform Spectrometer (FTS) with a spectral 
resolution of 50 kHz and 37275 channels. Two bands, 1.82 GHz wide each, were centered at 262.816 and 
265.886 GHz. These two bands were observed simultaneously in the two orthogonal polarizations of the
EMIR 230 GHz receiver.
All the 
observations were performed using the Wobbler Switching mode which produces remarkable flat
baselines. Pointing errors were always within 3$''$. The 30m beam size at the observing frequency
is 11$''$. The spectra were calibrated in antenna temperature corrected for atmospheric 
attenuation using the ATM package \citep{Cernicharo1985,Pardo2001}. 

\begin{figure}
\includegraphics[angle=0,scale=.6]{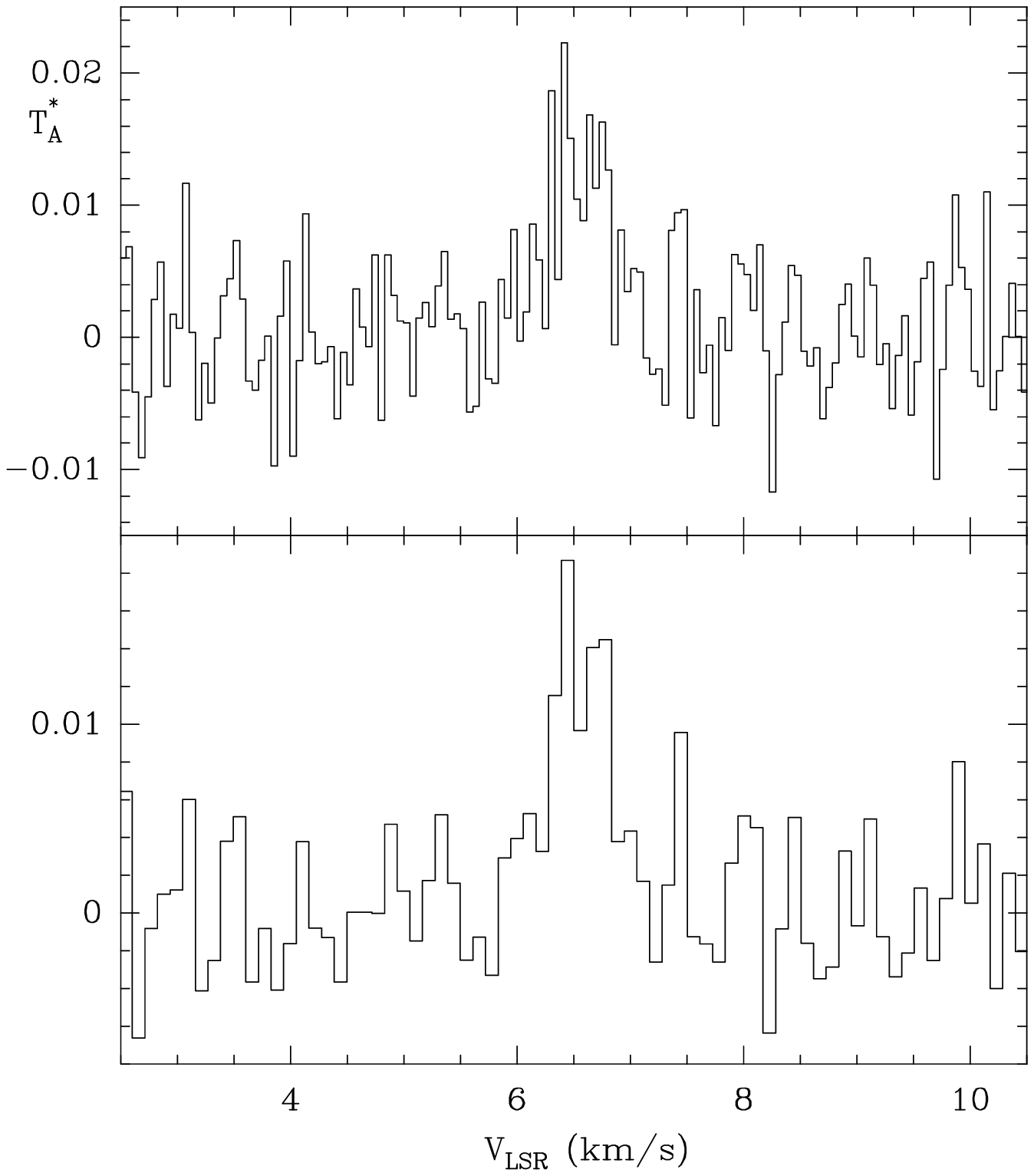}
\caption{Observed spectrum towards B1-bS ($\alpha_{2000}$=03$^h$ 33$^m$ 21.34$^s$,
$\delta_{2000}$=31$^o$ 07' 26.7'') at the expected frequency
of NH$_3$D$^+$ $J_K$=1$_0$-0$_0$ transition. The upper panel contains
the raw data resulting from a total integration time of 51 hours with a spectral
resolution of 48.8 KHz (0.054 km\,s$^{-1}$). The bottom panel shows the same data smoothed to 98 KHz.
For a LSR velocity of the source of 6.5 
km\,s$^{-1}$ the derived line frequency for the observed feature is 262816.73$\pm$0.1\,MHz.}
\end{figure}

Two different runs have been used for the present observations. The first one in July 2012 allowed us to
detect a feature at the same frequency of the line observed in Orion (262816.73\,MHz). During
this run weather conditions were mostly average to moderate summer, with 3--10 mm of precipitable water vapor. System 
Receiver temperatures were between 80 K and system temperatures always above 400 K. A second run in
January 2013 provided much better weather conditions and sensitivities. The water vapour amount 
was 1-2 mm and system temperatures between 200-250 K. The final spectrum, with all
data averaged, has a sensitivity of 4.9 mK with 48.9 KHz of spectral resolution (3.7 mK
after a box smoothing of 2 channels). The total observing time was 50.8 hours with an averaged
system temperature of 249 K.

\section{Results}
The resulting spectrum from the 30m telescope observations is shown in Figure 2. The upper panel
shows the data at the original spectral resolution of 48.9 KHz. The lower panel shows the same data
smoothed to 97.8 KHz. A feature is observed right at the expected frequency (see below) with a 
signal to noise ratio of
4.5 and 6 in the upper and lower panels, respectively. The line is narrow, having a width at half maximum of
0.65 km\,s$^{-1}$. It appears as having a double peak that could be due to the hyperfine structure but
it is within the noise of the data. Due to the central position of the nitrogen atom in Ammonium the electric
field on the nucleus is close to zero. Hence, the quadrupolar coupling constant of the nitrogen nucleus 
is probably negligible. The
Deuterium atom has a spin of 1 and could introduce also some hyperfine splitting but the corresponding
quadrupolar momentum is expected to be also very low. The identification relies on the close match 
between the frequency of the feature with that predicted from the rotational constants derived
with our new laboratory measurements (see below). 
Moreover, this is the unique feature detected together with the lines of CCH $N$=3-2 over a bandwidth
of 3.8 GHz. Taking into account the low kinetic temperature of the cloud no candidates can be found 
in MADEX, the CDMS, or JPL catalogues. Hence, the carrier has to be a light species and NH$_3$D$^+$ is 
the best candidate. Note that in this source NH$_3$, $^{15}$NH$_3$, NH$_2$D, $^{15}$NH$_2$D, ND$_2$H, 
ND$_3$ have been detected plus NNH$^+$, NND$^+$, $^{15}$NNH$^+$, N$^{15}$NH$^+$ (see above), 
so the chemistry is well 
constrained and the abundance of NH$_ 3$ and its related species is high.

\section{New Laboratory Experiments}
The density of lines in B1-bS is low at these high frequencies and the
agreement between the frequency of
the observed feature with the predictions of \cite{Nakanaga1986} is very good. However, 
a final
assignment relies on the precise determination from laboratory data of the rotational constants
of NH$_3$D$^+$. We have measured again the $\nu_4$ band of deuterated Ammonium with better absolute frequency 
calibration and more lines than those reported by \citet{Nakanaga1986} ($2\times10^{-4}$ vs. $1\times10^{-3}$ cm$^{-1}$ 
and 76 vs. 61 lines respectively). A detailed description of these experiments and calculations 
are presented in the accompanying Letter (Dom\'enech et al., 2013). 
The predicted frequency, calculated from the ground state rotational constants derived from a fit 
with the same parameters 
as those of \citet{Nakanaga1986} is 262817$\pm$6 MHz ($\pm 3\sigma$). 
The difference with the observed frequency is much less than one sigma deviation.
Hence, we are fully confident that the detected feature in Orion and B1-bS
corresponds to the 1$_0$-0$_0$ transition of NH$_3$D$^+$.

\section{Gas phase chemical modeling}
The Orion data will be interpreted in a forthcoming paper. In this work we will concentrate
on the analysis of the B1-bS data. 
Due to the low dipole moment of NH$_3$D$^+$, $\mu$=0.26 D, the Einstein coefficient for
spontaneous emission is A$_{ij}$=4.8$\times$ 10$^{-6}$ s$^{-1}$ and the critical density to efficiently
pump the 1$_0$ rotational level is just of a few 10$^4$ cm$^{-3}$. Hence,  
we could assume that the levels of NH$_3$D$^+$ are thermalized at $T_K$=12 K
\citep{Marcelino2005,Cernicharo2012}, we derive for the $A$ species of NH$_3$D$^+$ a
column density of (5.5$\pm$1.0)$\times$10$^{11}$ cm$^{-2}$. The total
column density of deuterated Ammonium is, hence, $N$(NH$_3$D$^+$)=(1.1$\pm$0.2$)\times$10$^{12}$ cm$^{-2}$.
The column density of molecular hydrogen derived by \citet{Hirano1999} is 
1.3$\times$10$^{23}$ cm$^{-2}$. 
Therefore, the abundance of deuterated Ammonium is $\simeq$8$\times$10$^{-12}$. 
NH$_2$D has been observed by \citet{Saito2000} who obtained a column density of (3.2$\pm$1.2)$\times$10$^{14}$
cm$^{-2}$. Hence, the abundance ratio of deuterated Ammonia over the deuterated Ammonium ion is
$\simeq$300. Assuming a deuteration
enhancement similar to that observed for NH$_2$D \citep{Saito2000} the abundance of
Ammonium is $\simeq$1.6\,10$^{-10}$.

All deuterated isotopologues of Ammonia have been detected in B1-bS \citep{Lis2002,Lis2006,Saito2000} and 
the corresponding gas phase chemical models have been presented by \cite{Roueff2005}
who were able to obtain satisfactory agreement between observations and gas phase models in 
which depletion effects of C and O on grains have been simulated.
In these models, Ammonia formation results directly from dissociative recombination of 
the deuterated Ammonium ion which gives preferentially the NH$_3$ + H product 
\citep{Ojekull2004}. As a consequence, these models suggest that Ammonium is present at a 
significative level in this environment and its possible detection is a major witness of 
the occurrence of such a gas phase chemistry.

Nitrogen chemistry has received considerable recent interest \citep{Dislaire2012} with 
the reevaluation of the starting step of gas phase chemical reactions leading to Ammonia 
via the Ammonium ion:  N$^+$ + H$_2$ $\to$  NH$^+$ + H.
This reaction is known to have a slight endothermicity  and its efficiency is very much
dependent on the para/ortho state of  H$_2$ i.e, wether rotational state of H$_2$ is 0/1, 
respectively. Indeed, based on low temperature experiments \citep{Marquette1985}, the 
endothermicity involved is 168.5 K with para-H$_2$ and 41.9 K when H$_2$ is in its ``normal'' 
form (1/4 para + 3/4 ortho). These values should be compared to the energy defect between 
the corresponding $J$=0 and $J$=1 levels of H$_2$, which is 170.5~K. \cite{Dislaire2012} 
have reevaluated the reaction rate constant involved with the ortho form of H$_2$
with 
significant differences compared to the previous estimate of \cite{Lebourlot1991}. 
An additional complexity, which is not yet fully understood, is the occurrence of fine structure 
of N$^+$, whose levels are also split by similar amounts. These effects are neglected up to now.

\begin{figure}
\includegraphics[angle=0,scale=.27]{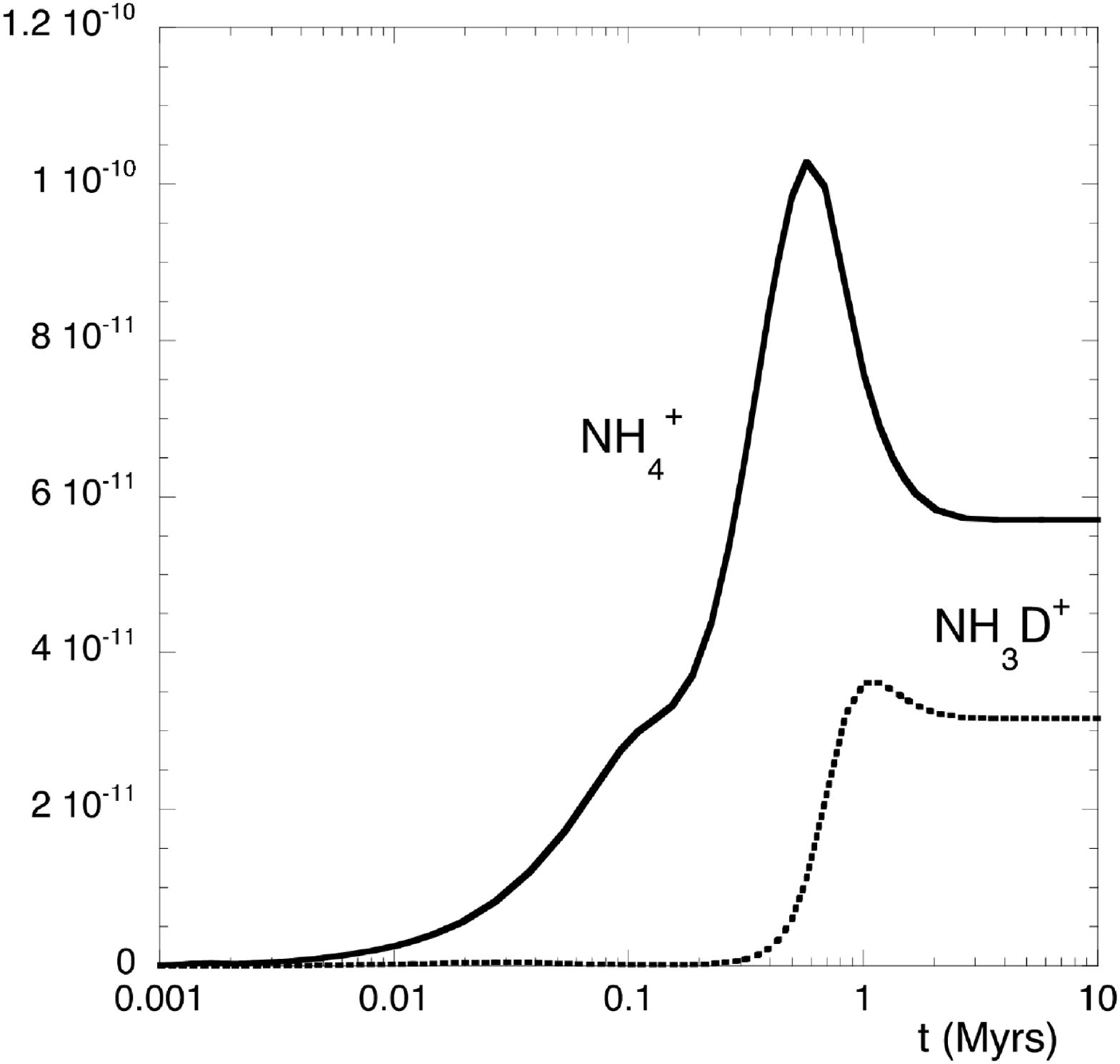}
\caption{Abundances of NH$_4^+$ and NH$_3$D$^+$ as a function of time for the chemical
model described in the text.}
\end{figure}

Interestingly enough, thanks to the lower zero-point vibrational energy of HD, the reactions

  N$^+$ + HD $\to$ ND$^+$ + H, 

  N$^+$ + HD $\to$ NH$^+$ + D 

\noindent
are found to have even smaller endothermicity.
We derive a value of 16.3\,K for the production of ND$^+$ and 100\,K for the channel towards NH$^+$. 
Once NH$^+$ or ND$^+$ are produced, they react with H$_2$, HD, D$_2$  in a sequence of rapid 
exothermic reactions until formation of Ammonium.
Walmsley and colleagues have been the first to include both para and ortho forms of H$_2$, related 
ions (H$_2^+$, H$_3^+$), and their multiply deuterated isotopologues for completely depleted regions 
\citep{Walmsley2004,Flower2004,Flower2006a,Flower2006b}, following an earlier suggestion of
\cite{Pagani1992}. We have extended this approach to a full chemical network including oxygen, 
carbon, sulfur and nitrogen \citep{Pagani2011}.
For the purpose of the present paper, we have updated this chemical network with the recent 
findings on nitrogen chemistry reported above on the one hand and included also some recently 
determined neutral-neutral reaction rate constants \citep{Daranlot2012}.
As a result, the present model contains 220 species (including ortho/para forms) of singly to 
fully deuterated substitutes and more than four thousand reactions.
We display in Figure 3 the corresponding time evolution of the fractional 
abundances relative to molecular hydrogen of NH$_4^+$ and NH$_3D^+$. The physical 
conditions correspond to those supposed to hold in B1-bS: the proton density is taken 
as 2$\times$10$^5$\,cm$^{-3}$, a temperature $T$=12\,K is assumed, and the elemental abundances 
correspond to model 2 of \cite{Roueff2005}. We see that the fractional 
abundances of NH$_4^+$ and NH$_3$D$^+$ peak are $\simeq$10$^{-10}$ (0.7\,Myr) 
and $\simeq$10$^{-11}$ (1\,Myr),
respectively which is consistent with the observational derivations. 
The abundance ratio $X$(NH$_4^+$)/$X$(NH$_3$D$^+$) is $\simeq$10. This value is similar
to that found for NH$_3$/NH$_2$D by \citet{Saito2000}. For t=1\,Myr this
ratio is $\simeq$2. 

Our chemical models predict that the abundance of NH$_2$D$_2^+$ 
could be a factor of two lower than that of NH$_3$D$^+$. Consequently, we could expect to detect  
more deuterated species of the Ammonium ion in cold dark clouds. However, more accurate
spectroscopic data are needed for NH$_2$D$_2^+$ and NHD$_3^+$ in order to search
for these species.


\acknowledgments
The Spanish authors thank the Spanish MICINN for funding support through grants, AYA2009-07304, CSD2009-00038, Fis2010-16455 and Fis2012-38175.
M. Gerin and E. Roueff acknowledge support from the french national PCMI program.
J. Cernicharo thanks U. Paris Est for an invited professor position during complection of this work.

\clearpage

\end{document}